# Electric Current and Noise in Long GaN Nanowires in the Space-Charge Limited Transport Regime

V. A. Sydoruk[*], S. A. Vitusevich[*], H. Hardtdegen[*,†]

[*]*Peter-Grünberg-Institute (PGI-8, PGI-9)*
*Forschungszentrum Jülich, Jülich 52425, Germany*

[†]*Jülich Aachen Research Alliance (JARA)*
*Jülich 52425 and Aachen 52062, Germany*

M. V. Petrychuk

*Taras Shevchenko National University of Kiev*
*Kiev 03022, Ukraine*

A. V. Naumov, V. V. Korotyeyev[‡], V. A. Kochelap[§]
and A. E. Belyaev

*Institute of Semiconductor Physics*
*National Academy of Sciences of Ukraine*
*Kiev 03028, Ukraine*

[‡]*koroteev@ukr.net*
[§]*kochelap@ukr.net*

We studied electric current and noise in planar GaN nanowires (NWs). The results obtained at low voltages provide us with estimates of the depletion effects in the NWs. For larger voltages, we observed the space-charge limited current (SCLC) effect. The onset of the effect clearly correlates with the NW width. For narrow NWs the mature SCLC regime was achieved. This effect has great impact on fluctuation characteristics of studied NWs. At low voltages, we found that the normalized noise level increases with decreasing NW width. In the SCLC regime, a further increase in the normalized noise intensity (up to $10^4$ times) was observed, as well as a change in the shape of the spectra with a tendency towards slope -3/2. We suggest that the features of the electric current and noise found in the NWs are of a general character and will have an impact on the development of NW-based devices.

---

[‡] Corresponding author

1. **Introduction**

Semiconductor nanowires (NWs) are attracting a great amount of attention because they demonstrate a number of new physical effects [1] and have good prospects for different applications in nanoelectronics, nanophotonics, sensing, etc. The use of different materials and techniques to fabricate single NWs and NW arrays allow observation of a variety of effects. Among these are effects which exist due to the geometry and the large aspect ratio of nanowires, rather than their material properties. In the simplest case, the geometry of a conductor can be characterized by the intercontact distance, $L$, and two dimensions transverse to the current, $W$ and $D$. If the aspect ratios $L/W$ and $L/D$ are large, the conductor can be thought of as a wire. The effects involving the electric charging of NWs are very different from those in conducting film structures ($W \gg L \gg D$) and bulk samples ($W, D \gg L$) due to a weak screening of the extra charges in NWs. The weak screening originates from the nonconductive dielectric surrounding [2].

In general, as the voltage, $U$, applied to a conductor increases, a linear Ohmic current regime is usually followed by a nonlinear current regime [3, 4]. If contacts to the conductor do not prevent charge injection, a superlinear current is observed when uncompensated injected charge carriers limit the current. For nanowires, charge injection effects are larger and occur at much lower voltages than in conductive film (F) and bulk (B) samples. Indeed, for trap-free samples of different dimensions, the space-charge limited currents (SCLCs) at large $U$ are estimated to be [3-6]

$$I^B = \frac{9}{32}\frac{\kappa\mu U^2}{\pi L}\frac{WD}{L^2}, I^F = \zeta_2 \frac{\kappa\mu U^2}{4\pi L}\frac{W}{L}, I^{NW} = \zeta_1 \frac{\kappa\mu U^2}{4\pi L}, \qquad (1)$$

where $\kappa$ is the dielectric constant of a bulk material, or that of the environment for film and NW systems, $\mu$ is the mobility; $\zeta_2$ and $\zeta_1$ are numerical coefficients of order of unity depending on contact geometry. Equations for $I^B$ and $I^F$ are valid for $(W \times D)/L^2 \gg 1$ and $W/L \gg 1$, respectively. From these equations, it follows that the currents are proportional to $U^2$ independent of the system dimensionality (the Mott-Gurney law). However, they have a considerably different dependence on the intercontact distance, $L$. A simple comparison of the ohmic currents with those given by Eq. (1) allows the estimation of the voltage when the system crosses over to the SCLC regime:

$$U_c^B = \frac{32\pi e}{9\kappa}n^B L^2, U_c^F = \frac{4\pi e}{\zeta_2 \kappa}n^B DL, U_c^{NW} = \frac{4\pi e}{\zeta_1 \kappa}n^B DW, \qquad (2)$$

where for convenience of comparison, we introduce an electron volumetric concentration, $n^B$. The above-mentioned inequalities imply that $U_c^B \gg U_c^F \gg U_c^{NW}$. Similar conclusions are also valid for samples with trapping centers, although to provide trap filling for such samples the currents increase as $U^{2+s}$ with $s > 0$ (see Refs. [4, 7]).

For nanowires, SCLC is a quite common phenomenon, which has been observed in nanowires grown and processed by different techniques and using different materials such as: GaN [6, 8], InAs [9], CdS [10], Si [11], and GaAs [12]. However, for better understanding of the electrical properties of nanowires, including electric current and noise and their exploitation in devices, investigations of the phenomenon are necessary taking into account basic characteristics of nanowires - dimensions, carrier mobility and concentration, contact properties, etc.

For NWs there is another common phenomenon related to charge carrier redistribution *across the wires*. Indeed, in the presence of surface states, for example deep defect states at the surface, the Fermi level is pinned at the surface causing a bending of the electron bands in the vicinity of the surface and forming a space charge redistribution. It is known, that for GaN NWs, the electronic bands are bent upwards toward the surface and an electron depletion region forms near the surface. The presence of both space charge redistribution phenomena significantly affects the electric current and noise of the NWs.

In our paper, we present the results of investigations of these phenomena in planar AlGaN/GaN nanowires, for which the basic characteristics are well defined. In addition, we study electric noise in the nonlinear transport regime, which has not been reported previously. The measurements of current noise, the constitutive aspect of the current penomena, revealed unusual electrical properties of the NW.

## 2. Nanowires under Investigation

We investigated multi-NW samples produced from planar AlGaN/GaN heterostructure wafers. The latter were grown by metal organic vapor phase epitaxy on a (0001) $Al_2O_3$ substrate. The heterostructure consisted of the 3 µm-thick GaN layer followed by a 40-nm-thick $Al_{0.1}Ga_{0.9}N$ top layer. The AlGaN/GaN NW structures were prepared by the use of electron beam lithography and $Ar^+$ ion beam etching. The etching depth of 95 nm was well below the depth of the AlGaN/GaN interface. A schematic cross section of the fabricated multi-NW samples is shown in the inset of Fig. 1. The method produced $N(=160)$ identical wires connected in parallel. Six sets of NWs of different widths and a fixed length of 620 µm were prepared and studied: $W$ = 280, 360, 470, 720, 930 and 1110 nm. On the same wafer, a 100- µm -wide Hall-bar structure was prepared. Ohmic contacts to NW and Hall-bar structures were formed by metal stack of Ti/Al/Ni/Au followed by rapid thermal annealing at 900°C for 30s. The current voltage (I-V) characteristics showed linear behavior at the voltages less than 0.1 V. The contact resistivity measured on the test structure was estimated as $2 \times 10^{-4}$ Ω $cm^2$. It was proved that for all NW structures the contact resistances were much smaller than the resistances of the NWs. Before proceeding with the experimental results, we should mention that in unintentionally doped AlGaN/GaN structures, built-in piezo- and polarization- fields form a quantum well for the electrons under the AlGaN/GaN interface. Even for a small Al content, the electrons are well confined near the interface, although their surface concentration, $n_{2D}$, is not large. For the heterostructure studied, temperature dependences of $n_{2D}$, and the mobility, $\mu(T)$ were measured, and are presented in Fig.1. Particularly, at room temperature we found $\mu = 1500$ $cm^2$/Vs and $n_{2D} = 2.1 \times 10^{12}$ $cm^{-2}$. We estimate the energy difference between the first excited and the ground subbands in the quantum well as 85 meV for the heterostructure studied. Thus at room temperature and below, the lowest subbands are predominantly populated.

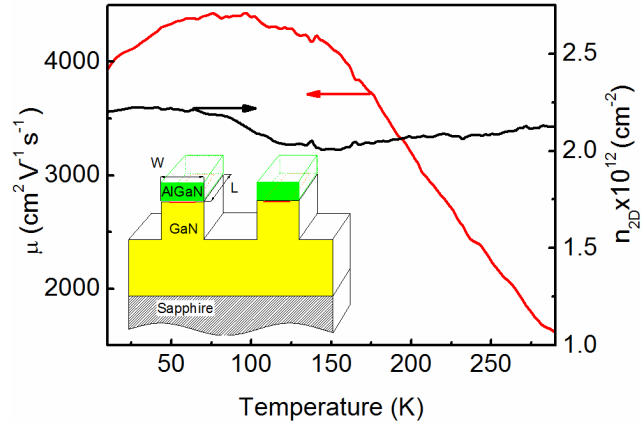

Fig. 1. The mobility and the surface electron concentration as functions of temperature for the AlGaN/GaN heterostructure studied. Inset: a schematic cross section of the heterostructure with parallel NWs. Current carrying nanowires are formed at AlGaN/GaN interfaces.

### 3. Current Regimes and Depletion Effect

We measured the current-voltage characteristics of all six sets of NWs. Some results are shown in Fig. 2. It can clearly be seen that in general, these characteristics are nonlinear. The analysis of the linear current regime leads to the conclusion that the conductance of NWs varies almost linearly with their width, as shown in the inset (a) of Fig. 2. Extrapolating these data, one can define the critical width, $W_c \approx 200$ nm, for which the conductance is expected to be zero. Indeed, for the structure with NWs of $185$ nm width the registered current was negligibly small. We also found that the critical width, $W_c$, is practically temperature independent. Moreover, the slopes of the fitting lines in inset (a) of Fig. 2 agree with the measured temperature dependence of the mobility $\mu(T)$.

The width dependence of the NW conductivity is explained by the formation of depletion zones near the edges of the NWs. Indeed, for GaN NWs, the depletion effect has been observed in experiments on conductivity [13], stationary and transient photoresponse [14, 15], and it has also been directly measured using Kelvin probe force microscopy [16]. The effect appears for NWs fabricated/grown by different methods. A general explanation of the effect is based on the surface Fermi level pinning within the band gap of GaN due to relatively deep surface defects. The surface (in our case the edges) becomes negatively charged forming an electron depletion region in the NW. By using this concept we calculated the depletion effect in the planar NWs under consideration. For this purpose, we solved the two-dimensional Poisson equation for the electrostatic potential, $\phi$,

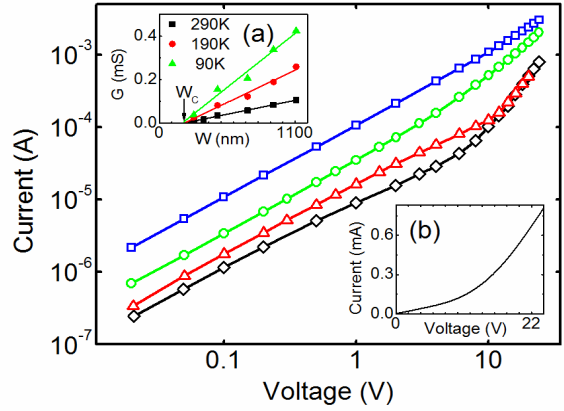

Fig. 2. The (I-V) characteristics for four sets of NWs (note the log-log scale), $W$=280, 360, 470, 1100 nm (currents are larger for wider nanowires). Inset (a): the low-voltage conductance vs the NW width at different temperatures. Inset (b): the I-V characteristics for the NW with $W$=280 nm. For the main figure and the inset (b), results are for $T$=290 K

self-consistently with the electron redistribution across the NWs. The problem was formulated in such a way that the only 'free' parameter was the one-dimensional density of the negative charge, $-eN^{edge}$, accumulated on an edge of the NW. The sketch of geometry of a NW sample with two-dimensional electrons and negatively charged edges and the frame of reference are shown in Fig. 3 (a).

The electrostatic problem is reduced to the solution of the coupled equations, which include: two-dimensional Poisson equation

$$\frac{\partial^2 (e\phi/k_BT)}{\partial X^2} + \frac{\partial^2 (e\phi/k_BT)}{\partial Z^2} = -2\pi B \delta[Z] \left[ 1 - \frac{n(X)}{N_D} - \frac{N^{edge}}{N_D W} \left\{ \delta\left(X - \frac{1}{2}\right) + \delta\left(X + \frac{1}{2}\right) \right\} \right] \quad (3)$$

the definition of the two-dimensional electron concentration obeying the Fermi statistics

$$n(X) = N_c \log\left[ 1 + \exp\left( \frac{E_F + e\phi(X,0)}{k_BT} \right) \right] \quad (4)$$

and the charge neutrality condition

$$N_D W = 2N^{edge} + W \int_{-1/2}^{1/2} n(X) dX \quad . \quad (5)$$

In Eqs. (3)-(5), we introduced the dimensionless coordinates normalized to the NW width, $X = x/W, Z = z/W$, $e$ is the elementary charge, $k_B$, and $\hbar$ are the Boltzmann constant and the Plank constant; respectively, $\kappa_0$ is the dielectric permittivity of a NW surroundings, $T$ is the temperature, $E_F$ is the Fermi level relative the bottom of the lowest subband, $\delta[Z]$ stands for the Dirac delta-function. $N_D$ is the concentration of the compensative positive charge (polarization charge or ionized donor charge), which is equal to $n_{2D}$ for original heterostructure, $B = 2e^2 N_D W / \kappa_0 k_B T$. Finally, $N_c = m^* k_B T / \pi \hbar^2$ is the density of states for the electrons in lowest subband with $m^*$ being the electron effective mass. Using the Green function for two-dimensional Poisson equation

$$G(X-X', Z-Z') = -\frac{1}{2\pi} \ln\left[\sqrt{(X-X')^2 + (Z-Z')^2}\right], \quad (6)$$

the formal solution of the Poisson equation can be written as follows:

$$\frac{e\phi(X,Z)}{k_B T} = \frac{B}{2}\left(\frac{N^{edge}}{N_D W} \ln\left[\left((X-1/2)^2 + Z^2\right)\left((X+1/2)^2 + Z^2\right)\right] - \int_{-1/2}^{1/2} dX'(1-n(X')/N_D) \ln[(X-X')^2 + Z^2]\right) \quad (7)$$

By using this equation, one can rewrite Eq. (4) in the form,

$$\frac{E_F}{k_B T} = \log(\exp(n(X)/N_c) - 1) - B\left[\log\left(\sqrt{|X^2 - 1/4|}\right)\int_{-1/2}^{1/2} dX'(1-n(X')/N_D) - \int_{-1/2}^{1/2} dX' \log(|X-X'|)(1-n(X')/N_D)\right] \quad (8)$$

This is the nonlinear integral equation for $n(X)$ with three parameters $N_c$, $B$, and $E_F/k_B T$. The first two parameters are given by the material properties of the NW. While ratio $E_F/k_B T$ is determined from the requirement that the solution of Eq. (8) should satisfy the neutrality condition of Eq. (5) at a given parameter $N^{edge}$. For realization of this procedure we applied the finite-element scheme to solve Eq. (8). As the results, we calculated the electron concentration $n(X)$ and the electrostatic potential $\phi(X,Z)$.

The parameter $N^{edge}$ was extracted from the conductance measurements presented in Fig. 2. Defining the conductance of the multi-nanowire structure, $G_0 = eN\mu n_{2D}W/L$, and using the measured conductance, G, we obtained the fraction of the electrons trapped at the edges, $\nu = (1 - G/G_0)$ and found $N^{edge} = \nu W n_{2D}/2$. The density of edge traps per unit NWs length is estimated as $2.2 \times 10^7$ cm$^{-1}$. For $W$ = 280, 360, 470 nm we found $\nu \approx 0.7$, 0.56, 0.42, respectively. Note, that $\nu = W_c/W$. For these parameters, distributions of the electron concentration and the dimensionless electrostatic energy (i.e., the electron band bending) across the NWs are presented in Fig. 3.

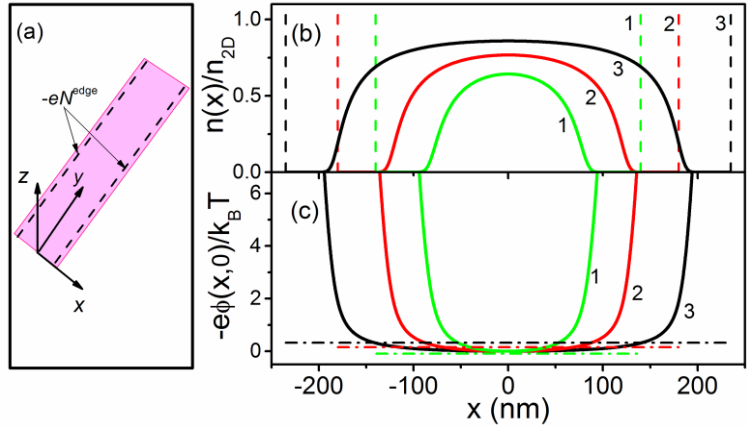

Fig.3. (a) Schematic geometry of a NW sample with two-dimensional electrons and negatively charged edges. (b) The distributions of the electron concentration across narrow NWs. The vertical dashed lines mark geometric sizes of the NWs; (c) The same for the electrostatic energy of the electrons (in units $k_BT$, at T=290 K). The dash-dotted lines indicate the respective Fermi levels. Curves 1, 2, 3 correspond to $W$=280, 360, 470 nm, respectively. $N_D = 2.1 \times 10^{12}$ cm$^{-2}$.

It is seen that at central parts of the NWs the electrostatic potential is flat and the electron concentration is a smooth function of the coordinate. Toward the edges, a fast increase in the potential energy (an upward band bending) creates completely depleted zones near the edges. The thicknesses of these depleted zones, with concentrations of less than $0.1 \times n_{2D}$, can be estimated as $W_d \approx$ 50..70 nm. Note, the total size of the depleted zones at both edges, $2W_d$, is less than the critical width, $W_c$, introduced above corresponding to almost zero conductance. This is because the middle parts of the NWs are partially depleted. This depletion effect is inherently large in narrow NWs. Indeed, for the examples presented in Fig. 3 the carrier concentrations reach maximal values $n_m/n_{2D} \approx 0.65, 0.76, 0.85$ at $W = 280, 360, 470$ nm respectively (the corresponding Fermi levels are small and change sign: $E_F/k_BT =$ -0.1, 0.16, 0.3).

Now we consider the nonlinear current regime. We begin this consideration with the remark that for the voltage interval studied ($U = 0..25$ V), the electric fields in our long NWs are small enough to neglect hot electron effects and related current nonlinearities [17]. From Fig. 2, it can be seen that the onset of the nonlinear regime depends on the wire width. For the voltage interval studied, this regime is most clearly apparent for the three narrowest NWs. At high voltage, two NWs exhibit super-Mott-Gurney dependences, which is characteristic of a developed regime of SCLC: $I^{NW} \propto U^{2.4}$ at $W = 280$ nm (shown in the inset (b) of Fig. 2), and $I^{NW} \propto U^{2.25}$ at $W = 360$ nm. For

*Electric current and noise in long GaN nanowires*

the NW of $W=470$ nm, we found $I^{NW} \propto U^{1.6}$. To understand such behavior, we note that in Eq. (2) for the critical voltage $U_c^{NW}$, the product $n^B DW$ means the number of electrons per unit wire length. For the planar NWs, this number can be estimated as $(1-\nu)n_{2D}W$. Using the parameters, $\nu$, above defined we calculated the critical voltages for the three NWs: $U_c \approx 3.7$, 5.5 and 12.5 V at $W=280$, 360 and 470 nm, respectively. Thus, for the first two NWs SCLC was achieved in our experiment, while for the third NW, only the regime transient to the SCLC transport was achieved. For the NW with 1100 nm width, we obtained $U_c \approx 40$ V, i.e., the SCLC regime was not achieved. This explains the current-voltage characteristics presented in Fig. 2. All features of the currents in the nonlinear regime were clearly observed while decreasing the temperature down to $T=90$ K.

**4. Electric Noise in Nanowires**

The studied features of electric currents in the NWs induce the new effect in electric fluctuations. We performned the investigations measurements of the spectral density of current noise, $S_I(f)$, in the frequency interval $f \leq 100$ kHz. The noise spectra were registered using a measurement system developed in-house and the Dynamic Signal Analyzer HP 35670A in the range from 1 Hz to 100 kHz. The intrinsic input-referred thermal noises of the preamplifier and ITHACO amplifier were measured as $2\times10^{-18}$ V$^2$/Hz and $2\times10^{-17}$ V$^2$/Hz, respectively. The detail description of the measurements can be found in Ref. [21]. In planar AlGaN/GaN heterostructures, the current noise has two main components: the flicker noise (with the Hooge parameter $\alpha_H \approx 10^{-4}...10^{-3}$) and generation-recombination (G-R) components, which appear when the temperature decreases [18-21]. Indeed, our noise measurements for the Hall-bar structure revealed flicker noise with a modest value of $\alpha_H \approx 2.9\text{x}10^{-3}$ and a few G-R components with activation energies close to those previously reported in Refs. [18, 19]. We found that compared to the planar heterostructures, the NWs are much more noisy. For *the linear current regime*, at $T=300$ K the flicker component ($\propto 1/f$) is dominant and the characteristic parameter $\alpha_H$ increases, when the NW width decreases: $\alpha_H \approx 8.7\text{x}10^{-3}$, 0.02 and 6.2, for $W=470$, 360, and 280 nm, respectively.

The high level of noise can be explained as follows. The low-frequency electric fluctuations arise due to processes of capture/decapture in defects. The presence of defects in extended AlGaN/GaN heterostructures determines a moderate flicker noise and the respective value of the Hooge parameter. In the NWs, there are additionally deep edge defects, which participate in both stationary phenomena, particularly determining depletion effects, and stochastic capturing/decapturing processes generating additional noise. The smaller the NW width, the larger the contribution of the edge defects to the noise, such that in the narrow NWs, the processes involving the edge defects are dominant.

For all NWs, the electric noise was also studied at *higher voltages*. Analysis of the measurements revealed two features of the noise generally found for all NWs: an increase in the noise intensity and a modification of the spectral density, $S_I(f)$, with increasing voltage/current. The noise increase with the voltage was stronger for NWs with lower *1/f* noise at small voltages. To illustrate this effect we analyzed the normalized spectral density, $S_I(f)/I^2$, as a function of the voltage for a given frequency, $f$. The results are presented in Fig. 4 for the NWs with $W = 360$ nm and 1100 nm. It can be seen, that the normalized noise spectra density increases by three to four orders of magnitude. For the NW with $W = 360$ nm, a steady noise increase appears in the voltage interval $U = 1...10$ V. For the wider NW, with $W = 1100$ nm, the noise intensity increases considerably at the larger voltages, $U > 10$ V. As the voltage increases, the spectral dependences, $S_I(f)$, of the NWs are modified. This is directly seen from the data presented in Fig. 4. Particularly, at $U \geq U_c$ the dependencies $S_I(f)$ show faster decay with $f$, as illustrated by the inset to Fig. 4 (a), for the narrow NW at the voltage $U = 8$ V. At the same voltage, the wider NW shows only the $1/f$ dependence (see the inset to Fig. 4 (b)).

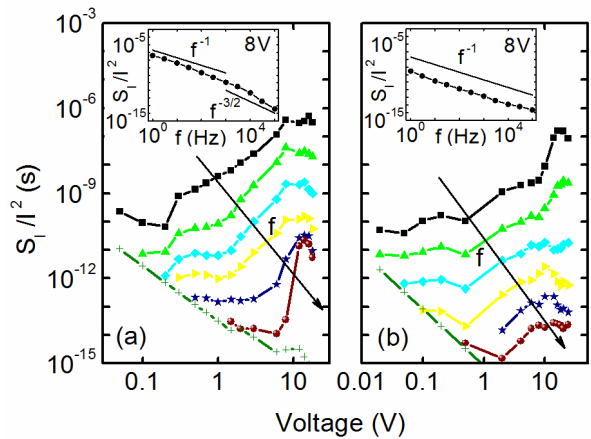

Fig.4. The normalized noise spectra density, $S_I(f)/I^2$, as a function of the applied voltage for f=1, 10, 100, $10^3$, $10^4$, $10^5$ Hz. The crosses indicate the estimated level of the thermal noise. In the insets: frequency dependences of $S_I(f)/I^2$ at $U=8$ V. (a): $W= 360$ nm; (b): $W= 1100$ nm.

The noise features of the NWs discussed above can be explained by the SCLC effects. The low-frequency noise is mainly determined by fluctuations of the free electron concentration. Under the conditions of the experiments presented here, the hot electron effects are negligible and distributions of the free electrons over the momenta and energies are almost in equilibrium. However, both the free electron concentration and that of electrons captured by the defects are far from equilibrium in the SCLC regime. Indeed, the main attributes of this regime are: electron injection, nonuniform electrostatic potential, development of a potential barrier near the cathode and redistribution of the electrons along the conductor [3-5]. Particularly, near the injecting electrode a zone with excess electron concentration is formed, while a zone near the anode is depleted due to exclusion processes. Obviously, both factors, excess and deficit of free electrons, give rise to intensified capture/decapture processes and to an increase of the noise intensity.

The additional mechanism of the fluctuations is provided by stochastic capture/decapture processes in the vicinity of the potential barrier: they randomly modulate the injection current and contribute to the noise.

Highly nonuniform electron distribution along the NWs in the SCLC regime explains the above mentioned modification of the noise spectral density, $S(f)$ at $U \geq U_c$. In fact, the $1/f$ noise dependence develops as the result of stochastic G-R processes with distributed characteristic times of capture/decapture events. This interpretation is valid for macroscopically uniform bulk samples [22] and for surface/edge noise in uniform conductive channels [22, 23]. However, for a nonuniform distribution of the electrons and, consequently, nonuniform noise sources, temporal evolution of a fluctuation involves diffusion processes. The latter results in so-called 'diffusion noise' [24], for which $S_I(f) \propto f^{-3/2}$. Note, this type of the noise is apparently different from the thermal Johnson-Nyquist current noise, intensity of which can be related to the diffusion coefficient. For NWs of length $L$, the characteristic diffusion time is estimated to be $\tau_D = L^2/D_0$, with $D_0 = k_B T \mu/e$ being the diffusion coefficient. The corresponding characteristic frequency is $f_D = 1/(2\pi\tau_D)$. For the experiments at $T = 300$ K, we found $D_0 \approx 37$ cm$^2$/s, $\tau_D \approx 10^{-4}$ s, $f_D \approx 1.5$ kHz This indicates that at $f > f_D$, the spectral density, $S(f)$, tends to be modified toward a $1/f^{3/2}$ dependence (see the inset to Fig. 4 (a)). For the wide NW (*W=1100* nm), the SCLC regime is not reached at studied voltages, the current regime still is linear, distributions of the carriers and noise sources are uniform. This results in the absence of the diffusion noise, as seen from the inset in Fig. 4 (b).

## 5. Conclusions

In conclusion, we studied electric current and noise in planar GaN NWs of different widths. The results obtained for low voltages allowed estimation of the depletion effects in the NWs. For the larger voltages, we found the current nonlinearities characteristic of the SCLC effect. The onset of this effect clearly correlates with the NW width. For the narrow NWs ($W = 280, 360$ nm), the muture SCLC regime was achieved. Noise effects studied in the NWs yielded additional proof for the entry into this regime. For low voltages, we found a large noise intensity for the narrow NWs. In the SCLC regime, a further increase in the noise intensity (up to $10^4$ times) was observed, as well as a change in the shape of the spectra, with a tendency towards a slope of -3/2. We suggest that the features of the electric current and noise in the NWs studied are of a general character. This knowledge can be important for the development of different NW-based devices.

**Acknowledgements**

This work is supported by the German Federal Ministry of Education and Research (BMBF Project 01DK13016). V. Korotyeyev and A. Naumov gratefully acknowledge for support FP7 project MCA-IRSES-294949 at Riga Photonics Centre of University of Latvia. Authors are grateful to Dr. Vanessa Maybeck for valuable discussions.